\pdfoutput=1
\documentclass[prb,aps,twocolumn,amsmath,amssymb,floatfix,superscriptaddress]{revtex4}
\usepackage[utf8]{inputenc}
\usepackage{textcomp}
\usepackage{color}
\usepackage{soul}
\usepackage{amsmath}
\usepackage{latexsym}
\usepackage{amssymb}
\usepackage{mathrsfs}
\usepackage{graphics,epstopdf}
\usepackage[colorlinks=true, citecolor=blue, urlcolor=blue ]{hyperref}
\usepackage{epsf,graphics,graphicx}
\usepackage{verbatim}

\date{\today}

\begin{document}
\title{Phases and phase transtions in one-dimensional alternating mixed spin ($\frac{1}{2}-1$) chain: effects of frustration and anisotropy}
\author{Soumya Satpathi}
\thanks{These authors contributed equally to this work.}
\author{Suparna Sarkar}
\thanks{These authors contributed equally to this work.}
\author{Swapan K. Pati}
\email{pati@jncasr.ac.in}
\affiliation
{Theoretical Sciences Unit, School of Advanced Materials (SAMat), Jawaharlal Nehru Centre for Advanced Scientific Research, Bangalore 560064, India}
\date{\today} 
\begin{abstract}
We investigate the phases and phase-transitions in one-dimensional alternating mixed-spin ($\tfrac{1}{2}$–$1$) chain in the presence of both frustration and anisotropy. Frustration is introduced via next-nearest-neighbor interactions, while single-ion anisotropy is incorporated at each lattice site. Our results show that moderate frustration can drive a phase transition from a ferrimagnetic state to an antiferromagnetic ground state. Remarkably, the presence of a weak easy-plane anisotropy destabilizes the ferrimagnetic order, also leading to the emergence of an antiferromagnetic phase. Interestingly, under strong frustration and anisotropy, the system exhibits signatures of a novel phase with spin density wave (SDW)-like modulation . We explore these anomalous phase transitions by employing exact diagonalization (ED) for small system sizes and the density matrix renormalization group (DMRG) method to characterize ground state properties for larger system sizes. We also investigate the finite-temperature behavior across various phases using the ancilla-based time-evolving block decimation (TEBD) approach. The primary objective of this work is to elucidate the phase structure of alternating mixed-spin chains under the combined effects of frustration and anisotropy. The primary objective of this work is to elucidate the intricate interplay between frustration and anisotropy in identifying the exotic phases and phase-transitions in alternating mixed-spin chains. Our findings contribute to a deeper understanding of mixed-spin quantum systems and may offer insights for future theoretical and experimental studies.
\end{abstract}

\maketitle

\section{Introduction}
Quantum spin chains constitute foundational models in the exploration of strongly correlated systems, providing critical insights into emergent phases of matter, quantum critical phenomena, and exotic excitations. In recent years, ferrimagnetic mixed spin chains with alternate spins, $s=\frac{1}{2}$ and $S>\frac{1}{2}$ have garnered a great deal of interest\cite{int1,int2,int3}. These systems display rich quantum behavior, most notably quantum phase transitions between distinct ground states, e.g. quantized magnetization plateaux\cite{int4} and Luttinger spin liquids\cite{int9}. Typically, quasi one-dimensional (quasi-1D) mixed-spin (MS) compounds exhibit antiferromagnetic (AFM) intrachain exchange interactions, which have spurred extensive theoretical investigations aimed at understanding their magnetic ground states. A wide variety of low-dimensional molecular magnetic structures containing different spins in a single unit cell have been successfully synthesized\cite{int5}. These systems exhibit novel quantum phases and unconventional thermodynamic properties. Most commonly, these systems consist of two transition metal ions within a single unit cell. Compounds, such as $NiCu(pba)(H_2O)_3$.$2H_2O$, characterized by alternating spins $(S_1,S_2) = (1,\frac{1}{2})$ and the family of $ACu(pbaOH)(H_2O)_3$.$nH_2O$, where A represents transition metal ions such as Ni, Co, Fe, or Mn—corresponding to spin pairs $(S_1,S_2) = (1,\frac{1}{2}),(\frac{3}{2},\frac{1}{2}),(2,\frac{1}{2}),(\frac{5}{2},\frac{1}{2})$, respectively. These serve as prototypical experimental platforms for realizing mixed-spin chains\cite{int5,swap2}. Similar mixed spin chain systems can also be experimentally realized in platforms of artificial quantum matter, including magnetic adatoms\cite{int6}, nanographenes\cite{int7}, and ultracold atoms  in optical lattices\cite{int8}.\par

Over the years models featuring competing nearest-neighbor (NN) and next-nearest-neighbor (NNN) interactions have extensively been studied \cite{int12,int13,int14}, particularly as a means to incorporate the effects of long-range Coulomb interactions. Alternating spin chains with NN Heisenberg interactions have been shown to exhibit  an AFM ground state and gapless ferromagnetic excitations. 
According to the Lieb-Mattis theorem\cite{intlsm}, the ground state of a Heisenberg ferrimagnet lies in the sector with total ground state spin 
$S = \frac{N}{2}(S_1 - S_2)$, where $N$ is the number of unit cells, and $S_1$ and $S_2$ are the alternating spin magnitudes. On a coarse-grained level, each spin pair (or dimer) behaves like a classical spin of magnitude $S = S_1 -S_2$, reflecting the net moment due to spin imbalance. However, within each dimer, quantum fluctuations persist\cite{swap1}. Typically, the inclusion of NNN coupling introduces frustration into the system, leading to a highly degenerate classical ground state\cite{int11}. However, this infinite degeneracy is lifted upon the introduction of quantum fluctuations, which select a host of interesting ground states, e.g., resonating valence bond, single-magnon states, spin-glass and spin-ice etc. Additionally, spin-orbit coupling and crystal-field interactions in magnetic ions generate preferred “easy” and “hard” axes for magnetization~\cite{int15}. Heisenberg chains with such single-ion anisotropy (SIA) have attracted considerable attention in both condensed matter and quantum information research~\cite{int151,int152,int153}. However, studies focusing on the effects of SIA in mixed-spin chains remain scarce. Experimentally, the SIA parameter, $D$ can be finely tuned through advances in nuclear electric resonance techniques\cite{int16,int17}. While alternating spin chains with NN AFM coupling have been the subject of extensive studies\cite{int9,swap3,swap1}, much less attention has been given to models that incorporate NNN AFM interactions and onsite anisotropy. Although understanding the nature of quantum phase transitions in low-dimensional systems remains a central focus in condensed matter physics\cite{int18,int19,int20}, the existence of quantum phase transitions in alternating spin chains remains less explored.

\par

\begin{figure}[ht]
	{\centering\resizebox*{8.5cm}{2.8cm}{\includegraphics{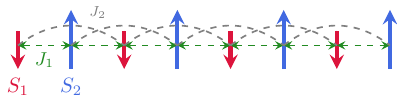}}}
	\caption{(Color online).  Schematic diagram of a spin $\frac{1}{2}$-spin $1$ alternate spin chain model, where nearest-neighbor interactions are $J_1$, and next-nearest-neighbor interactions are $J_2$. }
	\label{model}
\end{figure}
In this work, we report the emergence of a new quantum phase transition in the alternating Heisenberg spin chain composed of spins ($\frac{1}{2},1$), induced by a finite NNN AFM interaction. This transition marks a change in the ground state of the system from a commensurate ferrimagnetic to an incommensurate AFM phase. Additionally, we show that the presence of even a small easy-plane anisotropy destroys the ferrimagnetic order. Under conditions of strong frustration and large anisotropy, the originally ferrimagnetic system gives rise to a novel quantum phase, a low-dimensional spin-density wave phase in one of the spin sublattices. Here, we undertake a comprehensive investigation of this transition, focusing on the following key aspects: (i) the ground state properties to characterize each phase properly,
(ii) the nature of quantum phase transitions, and
(iii) the structure of low-energy excitations, as reflected in thermodynamic observables. To analyze these phenomena, we employ both numerical and analytical techniques. Exact diagonalization is used for studying small system sizes ($N \le 16$). To gain insight into the ground state and excitation spectrum in large systems, we further utilize perturbative linear spin-wave theory (LSWT) and the non-perturbative density matrix renormalization group (DMRG) method. Additionally, finite-temperature effects are explored using the time-evolving block decimation (TEBD) algorithm within the matrix product state (MPS) framework.

The structure of the remaining sections is organized as follows. In Section~II, we introduce the model Hamiltonian and outline the theoretical methods employed in this study. Section~III presents and discusses the numerical results in detail. Finally, the main conclusions are summarized in Section~IV.

\section{Numerical Methods} 

As illustrated in Fig. 1, we consider an alternating spin-($\frac{1}{2},1$) Heisenberg chain with NN and NNN AFM interactions. The Hamiltonian describing such a system with $N$ sites is given by

\begin{equation}\label{eq1}
H = J_1\sum_{i=1}^{N-1} S_{1,i}S_{2,i+1} + J_2\sum_{i=1}^{N-2}S_{1,i}S_{2,i+2},  
\end{equation}
where $J_1$ is the NN exchange coupling, and $J_2$ denotes the NNN exchange interaction. The operators $S_{1,i}$ and $S_{2,i}$ represent spin-$\frac{1}{2}$ and spin-$1$ operators on alternating lattice sites, respectively. To investigate the ground-state properties and quantum correlations in this system, we employ Fock space DMRG technique for system sizes upto ($\sim 100$), and a MPS-based DMRG approach for larger systems ($N \ge 100$), implemented using the ITensor library\cite{dmrg1,dmrg2,dmrg3}. The spin-spin correlation function is given by
\begin{equation}\label{eq2}
C^z(|i-j|) = \langle S^z_i S^z_j\rangle - \langle S^z_i\rangle - \langle S^z_j\rangle
\end{equation}
and the corresponding static structure factor as
\begin{equation}\label{eq21}
S(q) = \sum_{|i-j|}e^{iq|i-j|}C^z(|i-j|)
\end{equation}
The thermodynamic behavior of the system is thoroughly examined across a wide range of parameter regimes using the exact diagonalization (ED) approach. Specifically, we compute all eigenvalues of the model by diagonalizing it within fixed total magnetization sectors $M_s$, for a chain of $N \le 16$ sites. From these eigenvalues, we construct the canonical partition function $Z$ for the chain as
\begin{equation}\label{eq3}
Z = \sum_i e^{-\beta E_i}
\end{equation}
where the sum extends over all many-body eigen-states $i$, with $E_i$ denoting the energy and $M_s$ representing the $z$-component of the total spin for the $i^{th}$ state. Here, $ \beta/J_1$ equals to $1/k_BT$ with $J_1$ being the energy unit and $T$ the temperature. For convenience, temperature is measured in units of $k_B$ (considering $k_B=1$). The field-induced magnetization $M$ is then defined as the thermodynamic expectation value of the total spin along the field direction which is given by
\begin{equation}\label{eq4}
\langle M\rangle = \frac{\sum_i (M_s)_ie^{-\beta[E_i - (M_s)_i]}}{Z}
\end{equation}

The magnetic susceptibility, which quantifies the response of a system's magnetization to an applied magnetic field, can be defined in terms of the fluctuations in magnetization as
\begin{equation}\label{eq5}
\chi = \beta(\langle M^2 \rangle - \langle M \rangle^2)
\end{equation}
For larger systems, we employ the ancilla (or purification)\cite{nm1} approach to investigate finite-temperature properties. This method involves introducing an auxiliary set of fictitious states ${n'}$, each in one-to-one correspondence with the physical basis states ${n}$. In the enlarged Hilbert space, we define the unnormalized pure quantum state as
\begin{equation}\label{eq6}
|\psi(\beta)\rangle = e^{-\frac{\beta H}{2}}|\psi(0)\rangle = \sum_n e^{-\frac{\beta E_n}{2}}|n n'\rangle,
\end{equation}
where $\beta$ is the inverse temperature, and $|\psi(0)\rangle = \sum_n |n n'\rangle$ represents the thermal vacuum state at infinite temperature. The corresponding partition function is given by
\begin{equation}\label{eq7}
Z(\beta) = \langle \psi(\beta) | \psi(\beta) \rangle.
\end{equation}
The exact thermodynamic expectation value of an operator $A$, acting solely on the physical (real) degrees of freedom, is then calculated as
\begin{equation}\label{eq8}
\langle A \rangle = Z(\beta)^{-1} \langle \psi(\beta) | A | \psi(\beta) \rangle.
\end{equation}
We utilize the TEBD algorithm\cite{nm2,nm3,nm4}, as implemented in the ITensor library, to perform the imaginary-time evolution within the ancilla-based finite-temperature framework.

\section{Results} \label{sec:develop}
In this section, we present the key results derived from the theoretical framework discussed above. Our primary objective is to systematically examine the influence of frustration and anisotropy on the alternate spin $\frac{1}{2}$- spin $1$ chain and how the interplay between them leads to novel phases. We also discuss the effect of finite temperature on phase transition. Without loss of generality, we set the NN coupling to $J_1 = 1$ throughout the article unless otherwise stated. Our simulations are performed on chains with up to $N=240$ sites, using open boundary conditions and retaining up to $\sim800$ states in the renormalization process. The ground-state energy is well-converged, with numerical precision better than $10^{-7}$.
\subsection{Frustration}
\subsubsection{Spin wave analysis}
We start from the Hamiltonian in Eq.~\eqref{eq1}, describing a chain with alternating spins $S_1$ and $S_2$ on successive sites. The Holstein-Primakoff transformations
take the form
\begin{equation}\label{eq10}
\begin{split}
        & \hat{S}^z_{1,n} = S_1 - \hat{a}^\dagger_n\hat{a}_n,  \\
        & \hat{S}^+_{1,n} = \sqrt{2S_1-\hat{a}^\dagger_n\hat{a}_n} \hat{a}_n,\\
        & \hat{S}^-_{1,n} = 
        \hat{a}^\dagger_n \sqrt{2S_1-\hat{a}^\dagger_n\hat{a}_n}
    \end{split}
\end{equation}
for the spin-$S_1$ sites, and
\begin{equation}\label{eq11}
\begin{split}
        & \hat{S}^z_{2,n} = -S_2 + \hat{b}^\dagger_n\hat{b}_n,  \\
        & \hat{S}^+_{1,n} = \hat{b}^\dagger_n\sqrt{2S_2-\hat{b}^\dagger_n\hat{b}_n},\\
        & \hat{S}^-_{1,n} = 
        \sqrt{2S_2-\hat{b}^\dagger_n\hat{b}_n} \hat{b}_n
    \end{split}
\end{equation}
for the spin $S_2$ sites. 
Expanding the Hamiltonian to quadratic order and applying a Fourier transform, we diagonalize the resulting expression using a Bogoliubov transformation
\begin{equation}\label{eq12}
H = -2NJS_1S_2 + NJ_2(S_1^2+S^2_2) + \sum_k [\epsilon_{1k}c^\dagger_kc_k + \epsilon_{2k}d^\dagger_kd_k+\epsilon_{0k}]
\end{equation}
Here,
\begin{equation*}\label{eq13}
\begin{split}
   & c_k = a_k\cosh(\theta_k) + b^\dagger_{-k}\sinh(\theta_k) \\
   & d_k = b_{-k}\cosh(\theta_k) + a^\dagger_k \sinh(\theta_k)
\end{split}
\end{equation*}
and two distinct excitation branches with energies $\epsilon_{1k}$  and 
$\epsilon_{2k}$, along with a zero-point energy term $\epsilon_{0k}$ (for details, see Supplemental Material(SM).1~\cite{sup}), are explicitly given by
\begin{equation}\label{eq14}
\begin{split}
  &\epsilon_{1k} = (S_1 + S_2)[J_1 - J_2 + J_2\cos(k)] - \epsilon_k,\\
  &\epsilon_{2k} = (S_1 + S_2) [J_1 -J_2 +J_2\cos(k)] + \epsilon_k
\end{split}
\end{equation}
where
\begin{equation*}
\epsilon_k = \sqrt{(S_1-S_2)^2[J_1+J_2-J_2\cos(k)]^2+4J_1^2S_1S_2\cos^2(\frac{k}{2})}.
\end{equation*}

\begin{figure}[ht]
	{\centering\resizebox*{8cm}{4.5cm}{\includegraphics{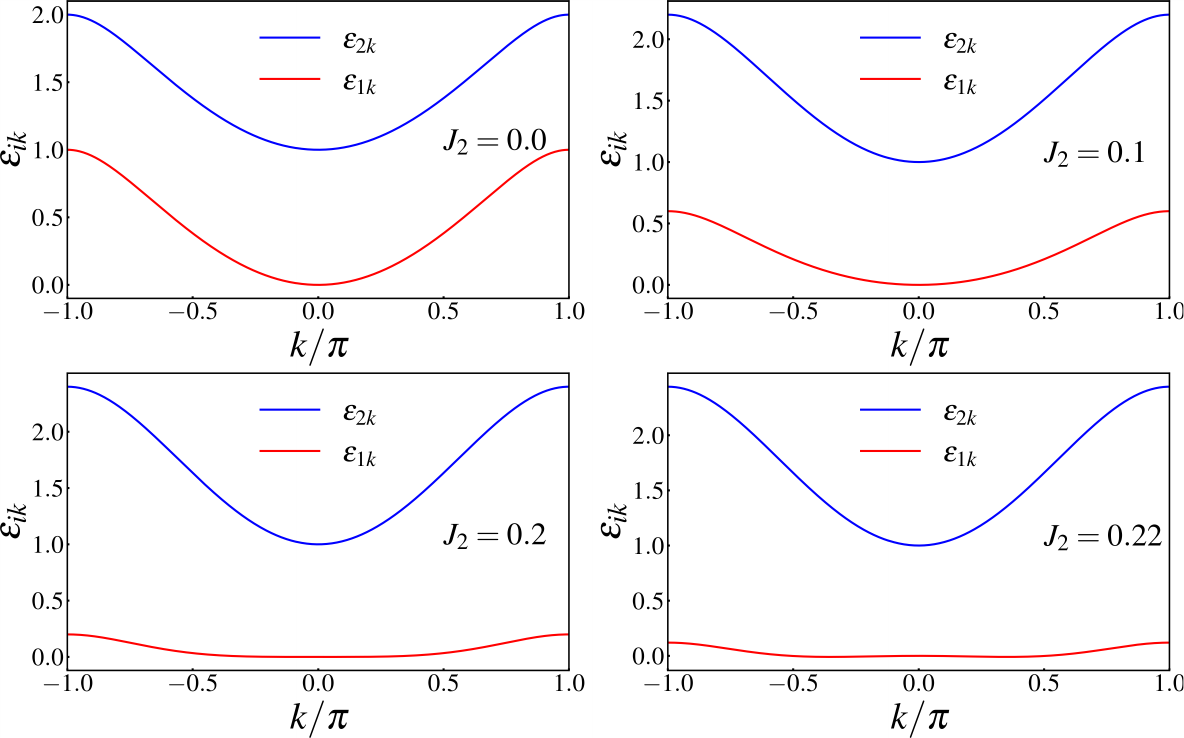}}}
	\caption{(Color online). The two spin-wave excitation branches, $\epsilon_{1k}$ and $\epsilon_{2k}$, expressed in units of $J_1$, are displayed for the alternating spin-($\frac{1}{2},1$) chain varying NNN coupling $J_2$.}
	\label{spinwave}
\end{figure}

The dispersion relations for the two excitation modes, $\epsilon_{1k}$ and $\epsilon_{2k}$ are presented in Fig. \ref{spinwave}. As evident from the figure, both $\epsilon_{1k}$ and $\epsilon_{2k}$ remain strictly positive across the entire momentum range from $k = -\pi$ to $\pi$. Consequently, the ground state corresponds to the state with $\langle c_k^\dagger c_k\rangle = \langle d_k^\dagger d_k\rangle = 0$. Analysis of the spin-wave energy spectrum reveals that the excitations associated with $\epsilon_{1k}$ are gapless at $k=0$, shown by red color curve. The energy spectrum corresponding to $\epsilon_{2k}$ are gapped for all values of $k$ as indicated by the blue curve, exhibiting a minimum energy gap of magnitude $2J_1(S_1-S_2)$ at $k=0$, valid for small value of the coupling constant $J_2$. 
At $k=0$, the energies of both excitation modes are independent of $J_2$, as the terms involving $J_2$ vanish due to $\cos(k) = 1$. This observation is consistent with the analytical expressions for $\epsilon_{1k}$ and $\epsilon_{2k}$ given in Eq.~\eqref{eq14}, which clearly show that at 
$k=0$, the energy depends solely on the NN exchange constant $J_1$. For finite $k$, however, the energies of both modes are influenced by both coupling parameters, $J_1$ and $J_2$. The energy associated with the $\epsilon_{1k}$ mode decreases with increasing $J_2$, suggesting that it softens under the influence of the NNN AFM interaction. This implies that the AFM coupling tends to suppress the underlying ferrimagnetic order in the system. Upon further increase of $J_2$, the dispersion of the AFM mode progressively flattens. At $J_2 \ge 0.23$, the mode energy becomes negative, and with continued increase, it eventually acquires complex values, signalling the breakdown of linear spin-wave theory.


\subsubsection{ED and DMRG Studies}
The alternate $S_1=\frac{1}{2}$ and $S_2=1$ spin-chain model, without the next-nearest neighbor, is free of frustration. The Lieb-Mattis theorem\cite{intlsm} ensures that the ground state of a chain with $N$ sites resides in the sector with total $S=\frac{N}{2}(S_1-S_2)$. The ground state is verified to belong to this spin sector for a chain comprising $N$ sites. In our system, we find that the ground state also lies in the spin sector with total spin $S=\frac{N}{2}(S_1 - S_2)$ through extensive exact diagonalization studies across a range of system sizes (up to $N\approx16$) for values of $J_2 < 0.23$. 
\begin{figure}[ht]
	{\centering\resizebox*{8.5cm}{4cm}{\includegraphics{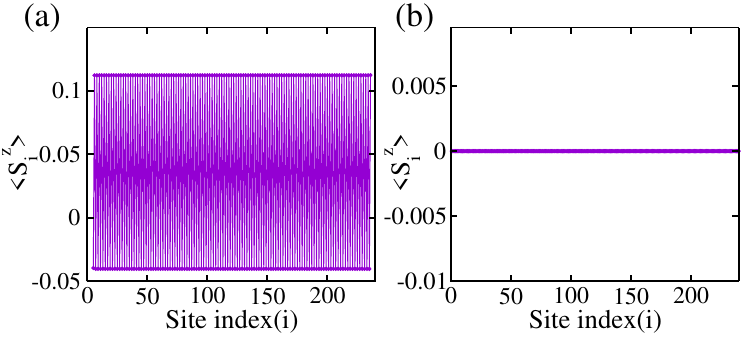}}}
	\caption{(Color online). Site-resolved expectation value of the $z$-component of the spin, $\langle S_z(i)\rangle$, for (a) $J_2=0.1$ and (b) $J_2=0.5$. }
	\label{spinden}
\end{figure}

Figure \ref{spinden} illustrates the site-resolved spin-density distribution along a chain of 240 sites for two representative values of the NNN exchange interaction, $J_2= 0.1$ and $J_2 = 0.5$. As shown in Fig. \ref{spinden}(a), for $J_2 = 0.1$, the spin-density remains uniform in each spin sublattice. Nevertheless, quantum fluctuations significantly renormalize the local spin expectation values: at the spin-$1$ sites, $\langle \hat{S}^z\rangle$ is reduced from the classical value of 1 to approximately $0.11$, while at the spin-$\frac{1}{2}$ sites, it is $-0.04$. Similar trend is observed in the ferrimagnetic chain even with only NN interactions due to quantum fluctuations\cite{swap1}. The spin-density profile reflects a ferrimagnetic-like alignment, where the net spin within each unit cell remains polarized with minimal variation across the sublattices. Interestingly, in the ferrimagnetic phase, the ground-state energy per site of the alternating spin system falls between those of the uniform spin-$\frac{1}{2}$ chain ($-0.443147J_1$) and the uniform spin-$1$ chain ($-1.401484J_1$), in agreement with earlier studies on NN alternating-spin chains\cite{swap1}.
For the NNN coupling $J_2=0.5$, the local spin densities at individual sites averaged out to zero as illustrated in Fig.~\ref{spinden}(b). This vanishing spin density signifies a transition to a nonmagnetic ground state, characteristic of a fully AFM phase. The observed behavior provides clear indication of a ferrimagnetic to AFM quantum phase transition in this alternating mixed spin chain.

\begin{figure}[ht]
	{\centering\resizebox*{8.5cm}{4cm}{\includegraphics{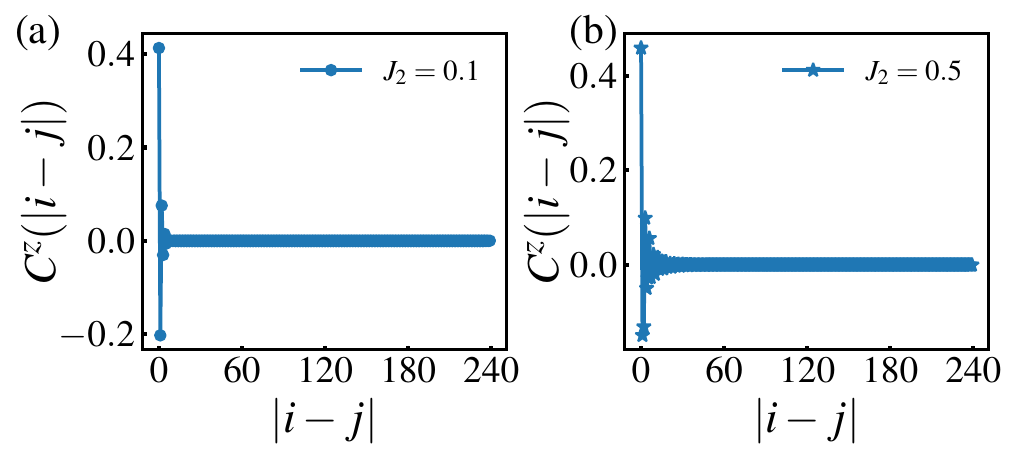}}}
	\caption{(Color online). Spin–spin correlation function $C^z(|i-j|)$ (defined in Eq.~\eqref{eq2}) as a function of the distance $|i-j|$ between two spins. (a) Corresponds to the phase with $J_2=0.1$, and (b) corresponds to the phase with $J_2=0.5$.}
	\label{spincorr}
\end{figure}

In Fig.\ref{spincorr}, the $C^z(|i-j|)$ correlation is plotted as a function of the distance between them for a system with chain length 240 sites. We also demonstrate the distinct types of spin-spin correlation functions ($\langle S^z_{1/2,1}S^z_{1/2,N}\rangle$, $\langle S^z_{1/2,1}S^z_{1,N}\rangle$ and $\langle S^z_{1,1}S^z_{1,N}\rangle$) that arise due to the alternation of spin-$\frac{1}{2}$ and spin-$1$ sites along the chain(see SM.2\cite{sup}). Here, Fig.\ref{spincorr}(a) exhibits a rapid decay of the static correlation as the individual spin averages remain finite. This behaviour is indicative of a magnetically ordered chain, specifically reflecting long-range order. In this case, the long-range order is driven by the presence of finite dimer magnetization across the lattice, characteristic of a ferrimagnetic ground state. However, at higher $J_2$, the $C^z(|i-j|)$ correlation function decays more slowly, as shown in Fig.\ref{spincorr}(b). The vanishing product of the individual spin averages indicates the absence of net magnetization, with spin singlet ground state and the gradual decay of correlations suggests the emergence of a quasi long-range order in the system.
\begin{figure}[ht]
	{\centering\resizebox*{8.5cm}{4cm}{\includegraphics{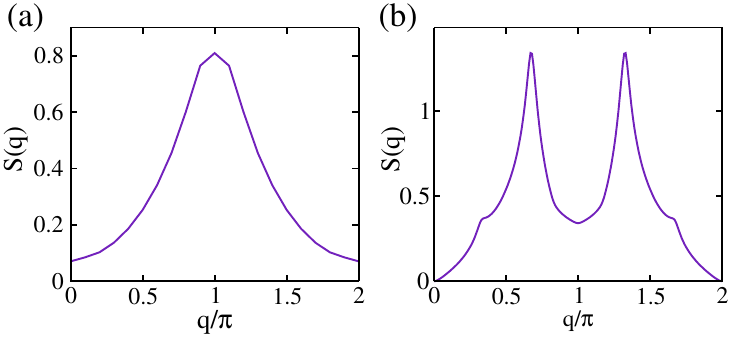}}}
	\caption{(Color online). The static spin structure factor $S(q)$ as a function of $q$ (defined in Eq.~\eqref{eq21}); (a) for $J_2 = 0.1$ and (b) for $J_2=0.5$.}
	\label{ssf}
\end{figure}

\begin{figure*}[ht]
	{\centering\resizebox*{7cm}{5cm}{\includegraphics{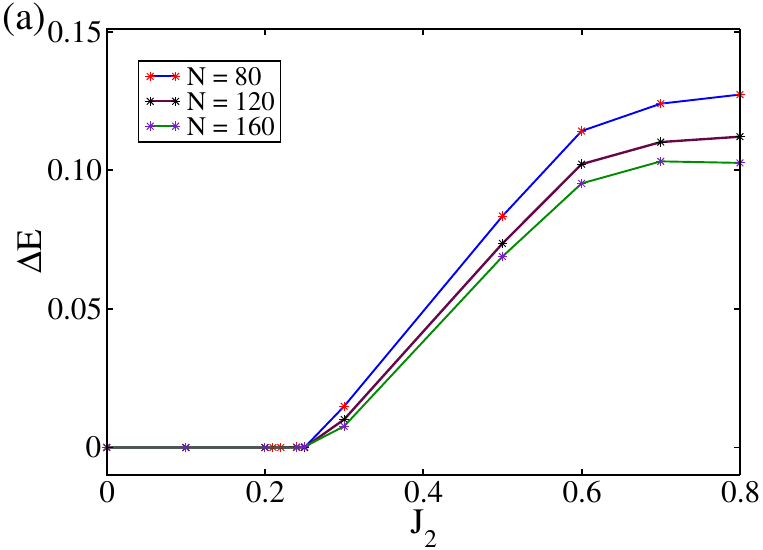}}
		\resizebox*{7cm}{5cm}{\includegraphics{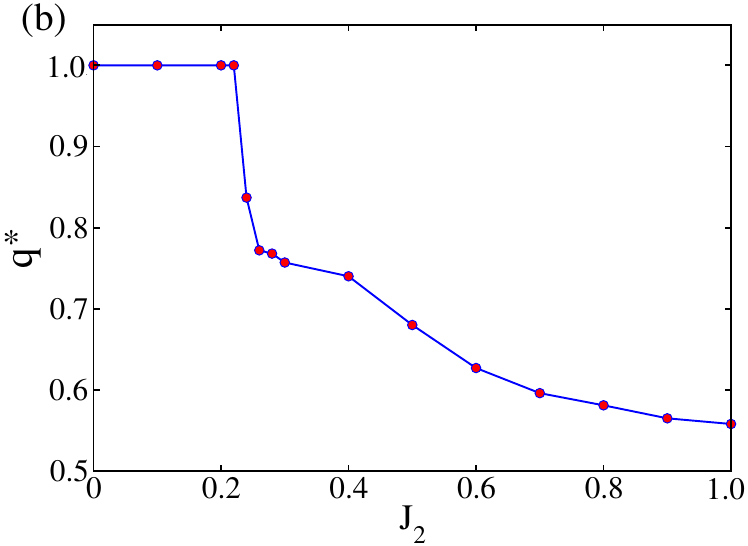}}
		\resizebox*{7cm}{5cm}{\includegraphics{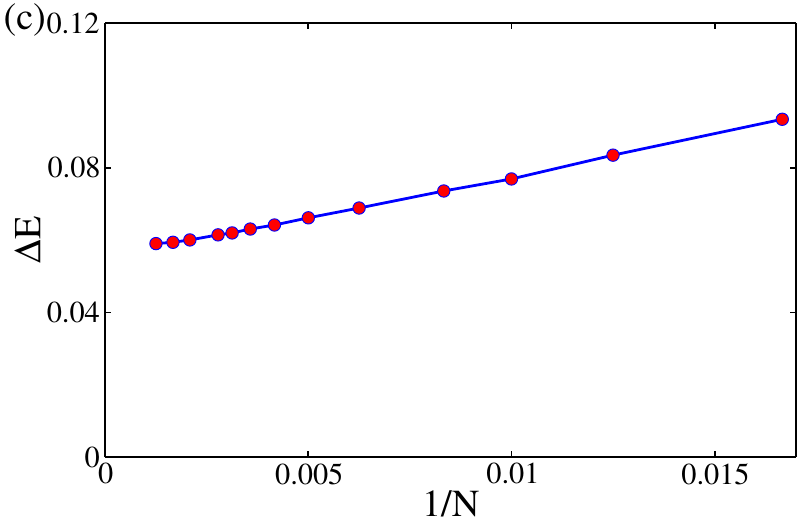}}
        \resizebox*{7cm}{5cm}{\includegraphics{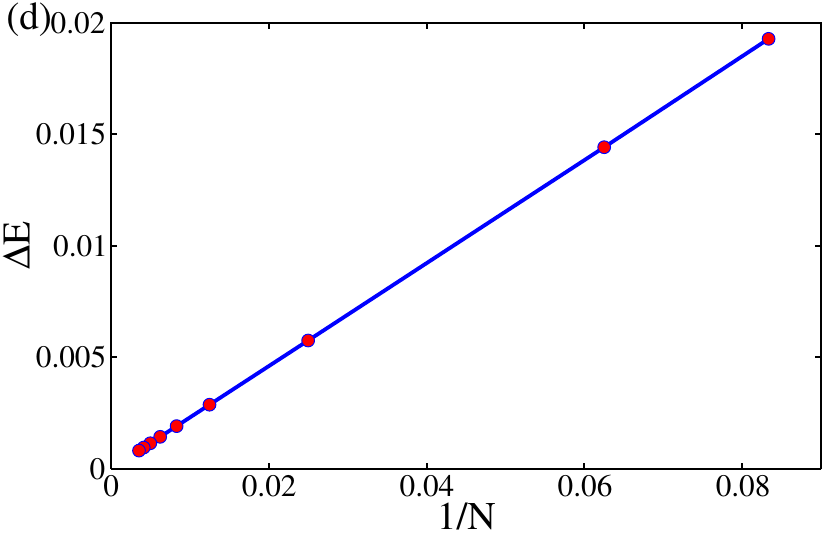}}}\par
	\caption{(Color online). (a) Singlet–triplet gap ($\Delta E$) for different system sizes as a function of NNN coupling $J_2$. (b) Variation of the peak position of the static structure factor (SSF) with $J_2$. (c) Finite-size scaling of the singlet–triplet gap ($\Delta E$) as a function of inverse system size ($1/N$) for $J_2=0.5$. (d) Finite-size scaling of the singlet–triplet gap ($\Delta E$) as a function of inverse system size ($1/N$) for $J_2=0.1, D=0.1$.}
	\label{gap}
\end{figure*}

Now, we plot the static spin structure factor (SSF) in Fig.~\ref{ssf} for two different values of $J_2$. For $J_2=0.1$, Fig.~\ref{ssf}(a) displays sharp peak at $q=\pi$, signifying well-defined magnetic order, which suggests the presence of commensurate (C) ferrimagnetic order. As $J_2$ increases shown in Fig.~\ref{ssf}(b), this peak at $q=\pi$ diminishes and eventually disappears, indicating the loss of a dominant ordering wavevector. Instead, two  peaks emerge at $\frac{\pi}{2}<q<\pi$ and $\pi<q<\frac{3\pi}{2}$, suggesting the onset of incommensurate (IC) magnetic correlations in the system.



\subsubsection{Gap analysis}
The critical value of $J_2$ corresponding to the transition from ferrimagnetic to antiferromagnetic order is identified by systematically analyzing the singlet-triplet excitation gap ($\Delta E$) using a combination of ED and DMRG calculations. Figure~\ref{gap}(a) shows the variation of $\Delta E$ as a function of $J_2$ for different system sizes. To further examine whether the system undergoes a C-IC transition at the same critical point, we tracked the momentum $q^*$ corresponding to the maximum of the SSF for various values of $J_2$, as shown in Fig.~\ref{gap}(b) (for details, see SM.4~\cite{sup}). We observe that the system evolves from a commensurate phase characterized by $q^* = 1$ to an incommensurate phase with $ 0.5 < q^* < 1.0 $ at $J_2 = 0.23$.
At this same value of $J_2$, Fig.~\ref{gap}(a) shows the opening of the singlet–triplet gap, signaling a transition from a gapless to a gapped phase. To assess whether the ferrimagnetic phase remains gapless in the thermodynamic limit, we analyze the lowest spin excitation corresponding to a transition from the ground state with total spin $S_G$ to a state with spin $S_G -1$. The excitation gap in the infinite chain limit was obtained by extrapolating the spin gap as a function of the inverse chain length (see SM.3~\cite{sup}). This analysis confirms that the ferrimagnetic phase is indeed gapless in the thermodynamic limit, with the ground state of total spin $S_G$ possessing ($2S_G+1$)-fold degeneracy corresponding to $S_G^z = -S_G,...,S_G$. A similar observation for the NN alternating spin chain was previously reported by Pati et al~\cite{swap1}.
In the AFM phase, we performed a comparable finite-size scaling of the excitation gap, as shown in Fig.~\ref{gap}(c). The extrapolation clearly indicates a very small yet finite gap in the $N\rightarrow \infty$ limit, with the ground state corresponding to a total spin $S_G=0$. These results collectively establish that the system undergoes a first-order quantum phase transition, characterized by the opening of a singlet–triplet gap as it evolves from a gapless ferrimagnetic phase to a gapped AFM phase. Consequently, the transition can be identified as a first-order C–IC quantum phase transition.

\subsection{Anisotropy}
\begin{figure*}[ht]
	{\centering\resizebox*{15cm}{8.5cm}{\includegraphics{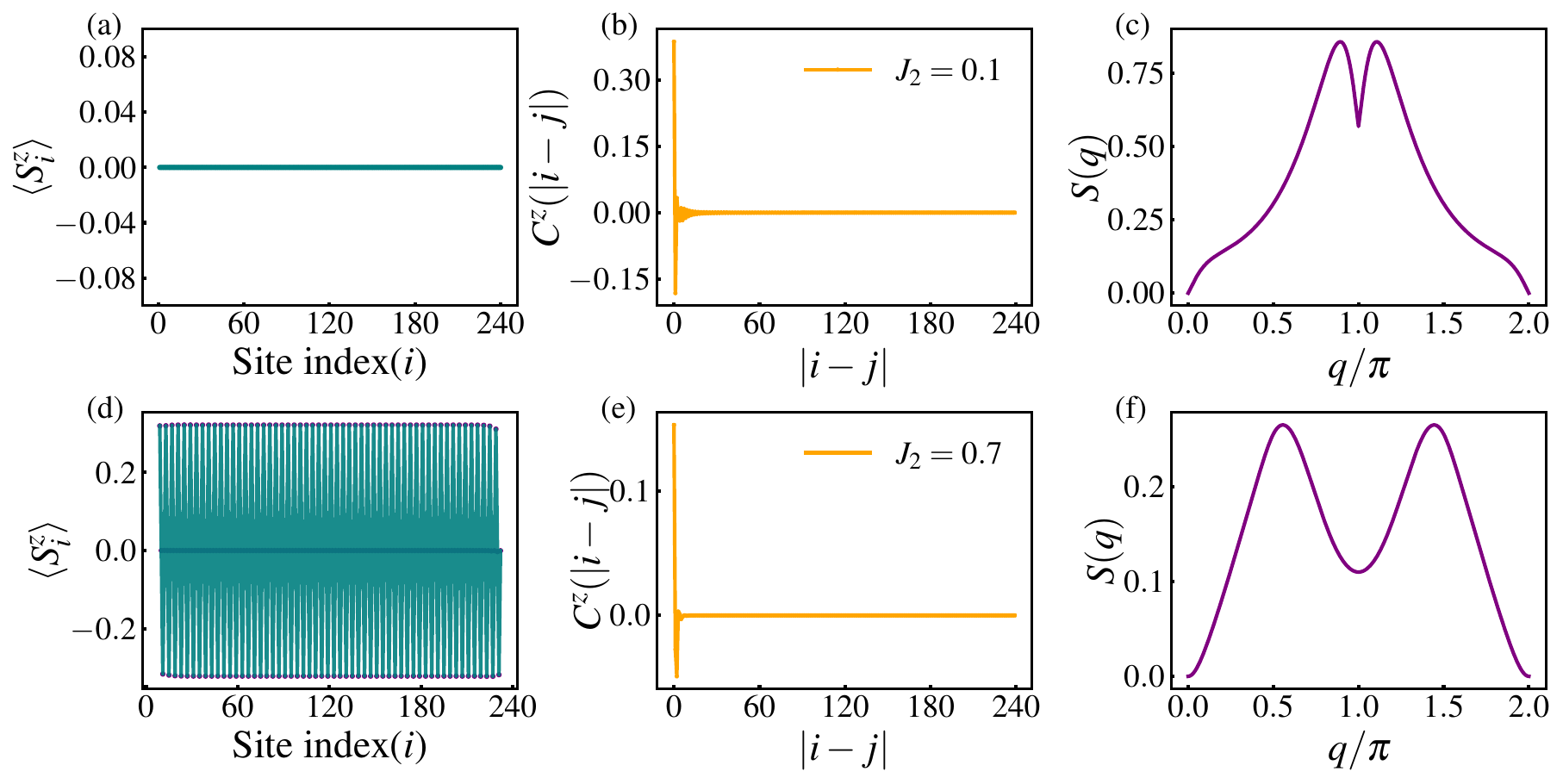}}}
	\caption{(Color online). (a,d) Expectation value of the $z$-component of the site spin at different sites, (b,e) spin-spin correlation as a function of distance between the two spins, (c,f) the static spin structure factor $S(q)$. Here, for, (a),(b),(c) $J_2=0.1, D=0.1$ and for (d),(e),(f) $J_2=0.7$ and $D = 2.0$.}
	\label{aniso}
\end{figure*}
We introduce single-ion anisotropy term as follows.
\begin{equation}\label{eq20}
H_{SIA} = D\sum_i(S^z_i)^2
\end{equation}
In the absence of anisotropy, the alternating spin $\frac{1}{2}$–spin $1$ chain remains in the ferrimagnetic phase for weak NNN coupling ($J_2 < 0.23$) and undergoes a transition to AFM phase at larger $J_2$. In this section, we investigate how the easy-plane SIA ($D>0$) influences these magnetic phases. Figures~\ref{aniso}(a)–(c) display, respectively, the site-resolved expectation value of the $z$-component of the spin, the spin–spin correlation function, and the static spin structure factor for $D=0.1$ within the ferrimagnetic regime ($J_2=0.1$). Furthermore, Figs.~\ref{aniso}(d)–(f) illustrate the corresponding quantities in the AFM phase, with strong frustration and enhanced anisotropy($J_2=0.7$ and $D=2.0$).\par
As shown in Fig.~\ref{aniso}(a), in contrast to the ferrimagnetic phase, the local spin densities at individual sites approach zero, indicating that even a small anisotropy (
$D=0.1$) drives the system toward a nonmagnetic ground state. The spin–spin correlation function in Fig.~\ref{aniso}(b) exhibits a slower decay compared to that of the ferrimagnetic phase, suggesting enhanced quantum fluctuations. We also present the distinct types of spin–spin correlation functions, ($\langle S^z_{1/2,1}S^z_{1/2,N}\rangle$, $\langle S^z_{1/2,1}S^z_{1,N}\rangle$ and $\langle S^z_{1,1}S^z_{1,N}\rangle$), which further confirm the suppression of long-range ferrimagnetic order and the emergence of a nonmagnetic, fluctuation-dominated ground state (see SM.5\cite{sup}). Moreover, the static structure factor, Fig.~\ref{aniso}(c), shows two pronounced peaks—one located between, $\frac{\pi}{2}$ and $\pi$ and the other between $\pi$ and $\frac{3\pi}{2}$. These results signaling the development of an incommensurate AFM ordering. Interestingly, unlike the frustration driven phase transition, the mixed spin chain does not develop a finite gap in this case, rather remains gapless in the thermodynamic limit. Figure~\ref{gap}(d) clearly demonstrates that the singlet–triplet gap diminishes with increasing system size and extrapolates to zero as $N\rightarrow \infty$. Hence, the C–IC phase transitions in these mixed spin chains driven by anisotropy are of a distinct nature.

In the AFM regime ($J_2>0.23$), the system exhibits a nonmagnetic ground state, as shown in Fig.~\ref{spinden}, where the local spin densities at individual sites average to zero. Interestingly, under strong frustration and large easy-plane anisotropy, the system evolves into a novel quantum phase in which only the spin-$\frac{1}{2}$ sites retain finite local spin densities, while the spin-1 sites display vanishing magnetization. As shown in Fig.~\ref{aniso}(d), the expectation value of $S_z$ on the spin-$\frac{1}{2}$ sites alternates between approximately $+0.32$ and 
$-0.32$ across neighboring sites, whereas the spin-$1$ sites exhibit negligible $S_z$ expectation values. The static spin structure factor, displayed in Fig.~\ref{aniso}(f), features a pronounced peak near 
$Q \approx 0.5$, accompanied by a rapid decay of the spin–spin correlations [Fig.~\ref{aniso}(e)]. This oscillatory behavior of the local magnetization, varying periodically ($\langle S_z^i\rangle \sim \cos(Q.i)$), indicates the emergence of a spin density wave (SDW), like modulation in the ground state.

\subsection{Finite Temperature}
\begin{figure}[ht]
	{\centering\resizebox*{4.2cm}{3.4cm}{\includegraphics{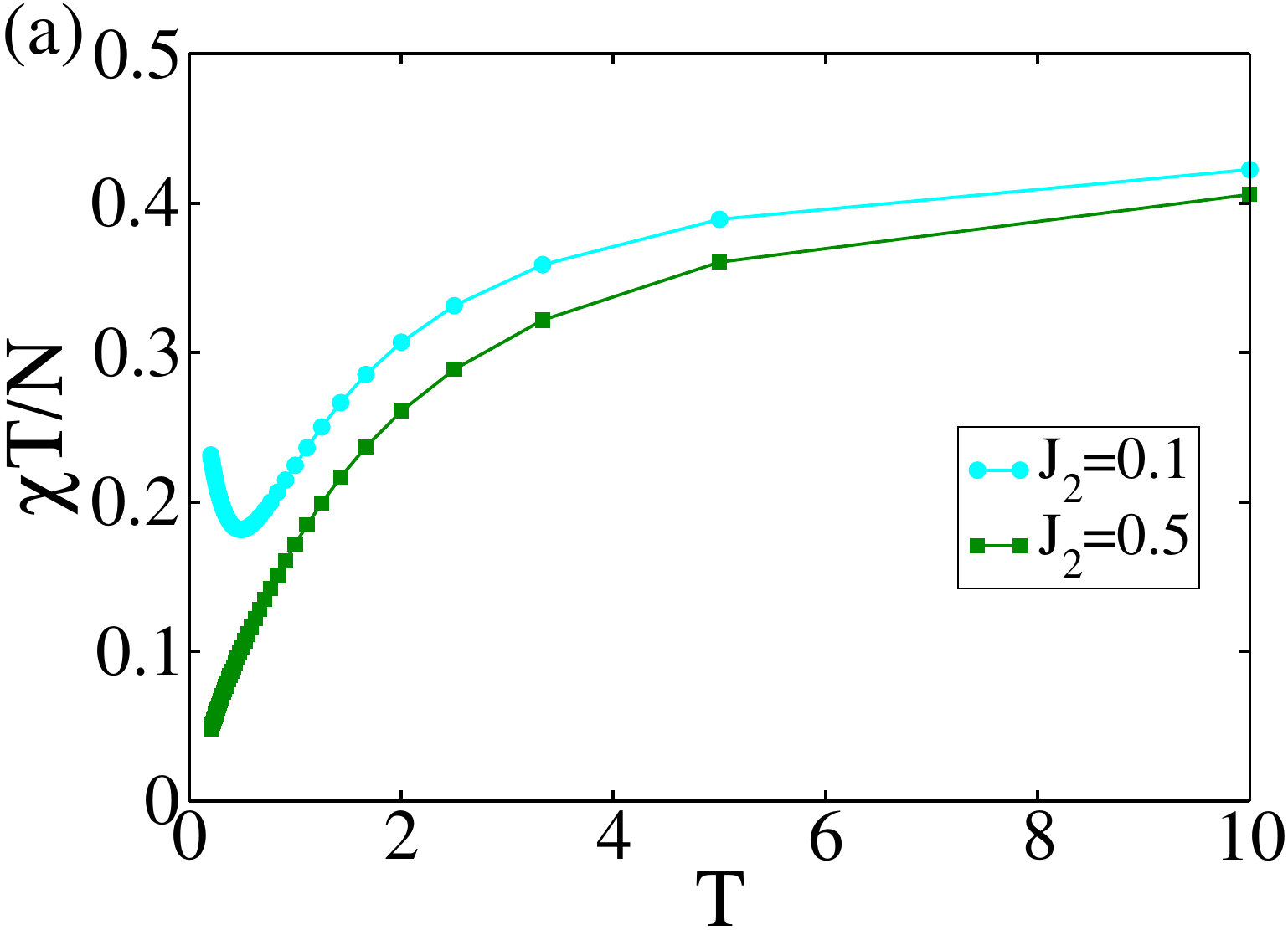}}}
	\resizebox*{4.2cm}{3.4cm}{\includegraphics{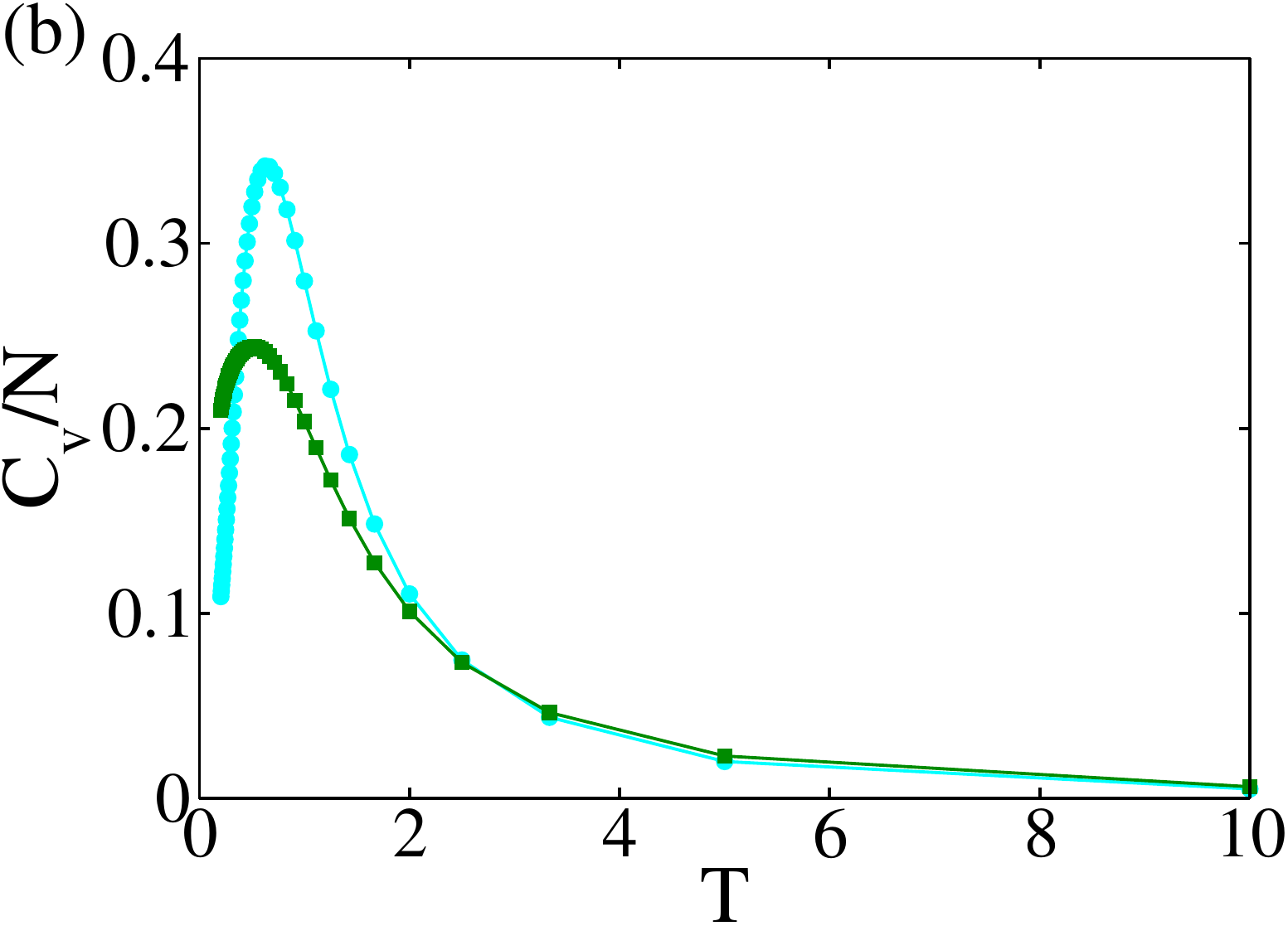}}\par
	\caption{(Color online). (a) Temperature dependence of the magnetic susceptibility product, $\chi T$, per site, and (b) specific heat, $C_v$, per site. The cyan curve corresponds to the ferrimagnetic phase with NNN coupling $J_2=0.1$, while the green curve represents the AFM phase with $J_2=0.5$.}
	\label{ss}
\end{figure}

To gain deeper insight into the finite-temperature behavior of the system, beyond the low-energy excitation regime, we compute its thermodynamic properties across different coupling regimes using ED and TEBD approaches. Specifically, we plot the temperature dependence of the magnetic susceptibility product, $\chi T$, per site and the specific heat, $C_v$, per site respectively in Fig.~\ref{ss}(a) and ~\ref{ss}(b). The cyan curve corresponds to the ferrimagnetic phase with NNN coupling $J_2=0.1$, while the green curve represents the AFM phase with $J_2=0.5$.
In the ferrimagnetic phase ($J_2=0.1$), shown in Fig.~\ref{ss}(a), as $T\rightarrow0$,  $\chi T/N = \frac{S_G(S_G+1)}{6N}$, as seen in previous studies also~\cite{swap1,swap3}. With increasing temperature, this quantity decreases and exhibits a minimum around $k_BT\sim0.5J_1$ before increasing again. The minimum arises due to the thermal population of low-lying states with total spin projections $M_s = S_G, S_G-1, ...,$ which correspond to the gapless excitations in the thermodynamic limit. The subsequent rise in $\chi T$ at higher temperatures results from the occupation of higher-energy, higher-spin states. In the AFM phase ($J_2 = 0.5$), $\chi T \rightarrow 0$ as $T\rightarrow 0$, consistent with a nonmagnetic singlet ground state. The specific heat, as shown in Fig.~\ref{ss}(b), further supports this behavior: the ferrimagnetic phase displays a higher-temperature peak, whereas the AFM phase shows a lower-temperature peak. These thermodynamic signatures confirm that the magnetic phases identified at zero temperature remain robust at finite temperatures.\par
To generalize our findings, we also investigate the effect of the aforementioned interactions on generalized spin-$\frac{1}{2}$-spin-$S$ chains. In particular, we consider two representative cases: the spin-$\frac{1}{2}$-spin-$\frac{3}{2}$chain and the spin-$\frac{1}{2}$-spin-$2$ chain. Our results indicate that the general spin-$\frac{1}{2}$–spin-$S$ chain similarly undergoes a quantum phase transition from a commensurate ferrimagnetic phase to an incommensurate AFM phase with increasing NNN coupling (for further details, see SM.6).



\section{summary and outlook}
In this study, we have investigated the ground state and thermodynamic properties of the alternating spin-$\frac{1}{2}$-spin-$1$ chain in the presence of both frustration and anisotropy. The system exhibits quantum phase transitions driven by the interplay of NNN frustration and SIA, which is also been studied thoroughly. Frustration is introduced via the NNN exchange coupling $J_2$, while anisotropy is incorporated through a SIA term of the form $DS_z^2$. We initially employed linear spin wave theory (LSWT) to gain insight into the excitation spectrum and phase behavior. However, as the frustration increases (i.e., with increasing $J_2$), quantum fluctuations become significant, leading to a breakdown of LSWT. Hence, we utilized ED techniques for small system sizes ($N \le 16$), and the DMRG method for larger chains, with system sizes up to $N \sim 240$. Calculations were performed for larger system sizes to reduce finite-size effects. To explore finite-temperature behavior, we employed the ancilla method incorporating TEBD technique to simulate the imaginary-time evolution. The main findings of our study are summarized below: \\
$\bullet$ The phases of a spin $\frac{1}{2}$- spin $1$ alternate spin chain with NNN coupling have been studied. The system exhibits a commensurate ferrimagnetic ground state at low $J_2$, which transits to an incommensurate AFM phase for $J_2 \ge 0.23$. \\ 
$\bullet$ The system undergoes a first-order C-IC quantum phase transition from the ferrimagnetic to the antiferromagnetic phase as a function of $J_2$. Similar phase transition is also observed for a spin $\frac{1}{2}$- spin $S$ chain. \\
$\bullet$ Using the ancilla approach, we confirm that the identified phases remain robust at finite temperatures.\\
$\bullet$ We introduce the easy plane anisotropy ($D>0$), which leads to suppression of ferrimagnetic order, destabilizing the commensurate ferrimagnetic phase. \\
$\bullet$ High anisotropy and high frustration lead to a novel phase of matter, where only the spin-$\frac{1}{2}$ sites have finite magnetization, and spin-$1$ sites has zero magnetization. The system behaves as a Neel state, considering each spin $\frac{1}{2}$- spin $1$ dimer, a site. This leads to a possible stable SDW phase in 1D.\\
Our results deepen the understanding of emergence of unconventional magnetic phases, such as incommensurate order and dimer-based SDW states in low-dimensional frustrated systems. The interplay of frustration and anisotropy in such mixed-spin chains may provide a fertile ground for realizing novel quantum materials and guiding experimental exploration in synthetic quantum magnets and cold atom systems.


\section*{ACKNOWLEDGMENTS}
SS acknowledges JNCASR, India for funding. SS is thankful to ANRF, India (File number: PDF/2023/000319) for providing her research fellowship. SKP acknowledges the JC Bose fellowship and SERB, Govt. of India for the financial assistance.


\begin{thebibliography}{99}
	
	
\bibitem{int1} W. M. da Silva and R. R. Montenegro-Filho, Magnetic-field–temperature phase diagram of alternating ferrimagnetic chains: Spin-wave theory from a fully polarized vacuum, Phys. Rev. B \textbf{96}, 214419 (2017).

\bibitem{int2}A. S. F. Tenório, R. R. Montenegro-Filho and M. D. Coutinho-Filho, Quantum phase transitions in alternating spin-($\frac{1}{2}-\frac{5}{2}$) Heisenberg chains, J. Phys.: Condens. Matter \textbf{23}, 506003 (2011).

\bibitem{int3}H. Yamaguchi, T. Okita, Y. Iwasaki, Y. Kono, N. Uemoto, Y. Hosokoshi, T. Kida, T. Kawakami, A. Matsuo and M. Hagiwara, Experimental realization of Lieb-Mattis plateau in a quantum spin chain, Scientific Reports \textbf{10}, 9193 (2020).   

\bibitem{int4}K. Maisinger, U. Schollw\"{o}ck, S. Brehmer, H. J. Mikeska and S. Yamamoto, Thermodynamics of the ($1,\frac{1}{2}$) ferrimagnet in finite magnetic fields, Phys. Rev. B \textbf{58}, 10 (1998).

\bibitem{int9} J. Stre\v{c}ka, Breakdown of a Magnetization Plateau in Ferrimagnetic
Mixed Spin-(1/2,S) Heisenberg Chains due to a Quantum
Phase Transition towards the Luttinger Spin Liquid, ACTA PHYSICA POLONICA A \textbf{131}, 264 (2017).

\bibitem{int5} J. S. Miller and M. Drillon, Magnetism: Molecules to Materials IV; John Wiley \& Sons: New York, NY, USA, 2002; p. 148.

\bibitem{swap2}S. Mohakud, S. K. Pati and S. Miyashita, Size-dependent low-energy excitations in an alternating spin-$1$/spin-$\frac{1}{2}$ antiferromagnetic chain: Spin-wave theory and density-matrix renormalization-group studies, Phys. Rev. B \textbf{76}, 014435 (2007).


\bibitem{int6} C. F. Hirjibehedin, C. P. Lutz, and A. J. Heinrich, Spin Coupling in Engineered Atomic Structures, Science \textbf{312}, 1024 (2006).

\bibitem{int7} S. Mishra, G. Catarina, F. Wu, R. Ortiz, D. Jacob, K. Eimre, J. Ma, C. A. Pignedoli, X. Feng, P. Ruffieux, J. F. Rossier and R. Fasel, Observation of fractional edge excitations in nanographene spin chains, Nature \textbf{598}, 292 (2021).

\bibitem{int8} P. Sompet, S. Hirthe, D. Bourgund, T. Chalopin, J. Bibo, J. Koepsell, P. Bojovi\'{e}, R. Verresen, F. Pollmann, G. Salomon, C. Gross, T. A. Hilker and I. Bloch, Realizing the symmetry-protected Haldane phase in Fermi–Hubbard ladders, Nature \textbf{606}, 488 (2022).



\bibitem{intlsm} E. Lieb, D. Mattis, Ordering Energy Levels of Interacting Spin Systems, J. Math. Phys. \textbf{3}, 751 (1962).

\bibitem{swap1} S. K. Pati, S. Ramasesha and D. Sen, Low-lying excited states and low-temperature properties of an alternating spin-$1$–spin-$\frac{1}{2}$ chain: A density-matrix renormalization-group study, Phys. Rev. B \textbf{55}, 14 (1997).

\bibitem{int11} H. J. Lee, M. Choi and G. S. Jeon, Emergent incommensurate correlations in frustrated ferromagnetic spin-$1$ chains, Phys. Rev. B \textbf{95}, 024424 (2017).

\bibitem{int12} A. Kolezhuk, R. Roth, and U. Schollw\"{o}ck, First Order Transition in the Frustrated Antiferromagnetic Heisenberg $S=1$ Quantum Spin Chain, Phys. Rev. Lett. \textbf{77}, 5142 (1996).

\bibitem{int13}  A. Kolezhuk, R. Roth, and U. Schollw\"{o}ck, Variational and density-matrix renormalization-group studies of the frustrated antiferromagnetic Heisenberg $S=1$ quantum spin chain, Phys. Rev. B \textbf{55}, 8928 (1997).

\bibitem{int14} J. H. Pixley, A. Shashi, and A. H. Nevidomskyy, Frustration and multicriticality in the antiferromagnetic spin-$1$ chain, Phys. Rev. B \textbf{90}, 214426 (2014).

\bibitem{int15} R. Morrow, A. E. Taylor, D. J. Singh, J. Xiong, S. Rodan, A. U. B. Wolter, S. Wurmehl, B. B\"{u}chner, M. B. Stone, A. I. Kolesnikov, A. A. Aczel, A. D. Christianson and P. M. Woodward, Scientific Reports \textbf{6}, 32462 (2016).

\bibitem{int151} Y. C. Tzeng and M. F. Yang, Scaling properties of fidelity in
the spin-$1$ anisotropic model, Phys. Rev. A \textbf{77}, 012311 (2008).

\bibitem{int152} S. Hu, B. Normand, X. Wang and L. Yu, Accurate determination of the Gaussian transition in spin-$1$ chains with single-ion anisotropy, Phys. Rev. B \textbf{84}, 220402 (2011).

\bibitem{int153} J. Ren, Y. Wang and W. L. You, Quantum phase transitions in
spin-$1$ XXZ chains with rhombic single-ion anisotropy, Phys. Rev. A \textbf{97}, 042318 (2018).

\bibitem{int16} S. Asaad, V. Mourik, B. Joecker, M. A. I. Johnson, A. D. Baczewski, H. R. Firgau, M. T. Madzik, V. Schmitt, J. J. Pla, F. E. Hudson, K. M. Itoh, J. C. McCallum, A. S. Dzurak, A. Laucht and A. Morello, Coherent electrical control of a single high-spin nucleus in silicon, Nature \textbf{579}, 205 (2020). 

\bibitem{int17} W. Lin, Y. Xu, Z. Liu, C. Wang and X. Kong, Single-ion anisotropy effects on the critical behaviors of quantum entanglement and correlation in the spin-$1$ Heisenberg chain, J. Phys.: Condens. Matter \textbf{33}, 345802 (2021). 

\bibitem{swap3} S. K. Pati, Alternating spin-$1$/spin-$\frac{1}{2}$ sites in a diamond lattice: Ground state and excitations, Phys. Rev. B \textbf{67}, 184411 (2003).

\bibitem{int18} S. Sachdev, Quantum Phase Transitions (Cambridge University Press, Cambridge, England, 2011) 

\bibitem{int19}	T. Giamarchi, Quantum Physics in One Ddimension, Internatational Series of Monographs on Physics (Clarendon, Oxford, 2004).

\bibitem{int20} N. Chepiga, From Kosterlitz-Thouless to Pokrovsky-Talapov transitions in spinless fermions and spin chains with next-nearest-neighbor interactions, Phys. Rev. Research \textbf{4}, 043225 (2022).

\bibitem{dmrg1} M. Fishman, S. R. White and E. M. Stoudenmire, The ITensor Software Library for Tensor Network Calculations, SciPost Phys. Codebases, 4 (2022).

\bibitem{dmrg2} M. Fishman, S. R. White and E. M. Stoudenmire, Codebase release 0.3 for ITensor, SciPost Phys. Codebases, 4-r0.3 (2022).

\bibitem{dmrg3} ITensor Library, http://itensor.org.

\bibitem{nm1} A. E. Feiguin and S. R. White, Finite-temperature density matrix renormalization using an enlarged Hilbert space, Phys. Rev. B \textbf{72}, 220401 (2005).

\bibitem{nm2} U. Schollw\"{o}ck, The density-matrix renormalization group in the age of matrix product states, Annals of Physics \textbf{326}, 96 (2011).

\bibitem{nm3} S. R. White, Density matrix formulation for quantum renormalization groups, Phys. Rev. Lett. \textbf{69}, 2863 (1992).

\bibitem{nm4} S. R. White, Density-matrix algorithms for quantum renormalization groups, Phys. Rev. B \textbf{48}, 10345 (1993).

\bibitem{sup} Supplemental Material




\end{thebibliography}
\end{document}